\documentclass[showpacs,prb,twocolumn,superscriptaddress]{revtex4}
\usepackage{epsfig}
\usepackage{bm}

\begin{document}

\title{An effective spin-orbital Hamiltonian for the double perovskite Sr$_2$FeWO$_6$:\\ Derivation of the phase diagram. }
\author{S. Di Matteo}

\affiliation{Laboratori Nazionali di Frascati, INFN, Casella Postale 13, I-00044 Frascati, Italy}
\affiliation{INFM UdR Roma III, via della Vasca Navale, 00100 Roma, Italy}
\author{G. Jackeli}
\affiliation{Institut Laue Langevin, B. P. 156, F-38042,
 Grenoble, France}
\author{N. B. Perkins}
\affiliation{Laboratori Nazionali di Frascati, INFN, Casella Postale 13, I-00044 Frascati, Italy}
\affiliation{BLTP, JINR, Dubna, 141980, Russia}
\date{\today}
\begin{abstract}

We formulate a superexchange theory of insulating 
double-perovskite compounds such as Sr$_2$FeWO$_6$.
An effective spin-orbital Hamiltonian is derived in the strong coupling 
limit of Hubbard model for d-electrons on Fe and W ions. The relevant degrees of freedom are the spins $S=2$ 
 and the three-fold orbital degeneracy of Fe$^{2+}$-ions. W-sites are integrated out by means of a fourth-order 
perturbative expansion. 
The magnetically and orbitally ordered ground states of the effective Hamiltonian are discussed as a function of the model parameters.
We show that  for realistic values of such parameters the ground state is 
antiferromagnetic, as experimentally observed. 
The order found is of type-II,  consisting  of \{111\} ferromagnetic planes stacked antiferromagnetically. 
The orbital order energy scale found is one order of magnitude less than the spin one.
\end{abstract}
\pacs{75.10.-b, 75.30.Et, 75.50.Ee}
\maketitle

\section{Introduction.}
Transition metal oxides have been the subject of numerous experimental and theoretical studies after the discovery of a variety of novel physical phenomena and a diversity of ordered phases.\cite{imada}
Their ground state can be insulating, metallic or superconducting, depending on temperature and stoichiometry 
and their magnetic structure varies from ferromagnetic (FM) to different antiferromagnetic 
(AFM) types. The insulating behavior is generally attributed
to a large Coulomb interaction that localizes the $d$-electrons in Mott-Hubbard
or charge-transfer regimes.\cite{imada} In this case
low energy excitations can be described by the spin degrees of freedom through 
an effective superexchange model. Often, the ground state of the transition metal ion  possesses an orbital degeneracy in addition to its that originating from the spin. In such cases, orbital degrees of freedom are also incorporated 
in the superexchange theory and the system is described by means of an effective
spin-orbital model.\cite{kugel}

The purpose of the present work is to formulate such a superexchange theory  and derive a spin-orbital Hamiltonian for the antiferromagnetic insulator 
Sr$_2$FeWO$_6$. The experimental value of its N\`{e}el temperature $T_N$ is quite low and reported in the literature as $16-37$ K.\cite{blasse,kawanaka,kobayashi1} 
Sr$_2$FeWO$_6$ belongs to a family of 
double perovskite compounds with general formula A$_2$BB$^\prime$O$_6$. It has,  
however, electronic and magnetic properties 
which differ significantly from those of the other members of the family.
For example, Sr$_2$FeMoO$_6$ and  Sr$_2$FeReO$_6$, that have been the subject of the intense studies 
since the discovery of room temperature magnetoresistance in these compounds,\cite{kobayashi2}   are 
 half-metallic ferrimagnets with fairly high magnetic transition temperatures.
\cite{kobayashi2}
One of the reason why  Sr$_2$FeWO$_6$ is so different from  the other members of the group has been suggested by band structure analysis.\cite{fang} 
In this picture, the antibonding W(5d) states are pushed  higher in energy
by the stronger hybridization with oxygen $p$-orbitals.\cite{fang,solovyev}
(In the case of Re(5d) states this effect is compensated by the deeper $5d$-level 
of Re ion.)  Because of this, 5d-electron prefer to move away from W site, leaving it in a W$^{6+}$ ($5d^0$) configuration, and stay on the Fe $3d$-level even at the cost of paying an extra Coulomb energy. As a result, in Sr$_2$FeWO$_6$ iron 
is in a Fe$^{2+}$ ($3d^6$)  valence state,\cite{dass} 
 and the moderately large Fe-W charge transfer gap, in the presence
of strong Coulomb interaction on Fe sites,  causes the insulating behavior.

In the perovskite structure, the octahedral ligand field of oxygens 
splits d-levels into a low-lying triplet
($t_{2g}$) and a higher doublet ($e_g$). 
There are several experimental evidences\cite{kawanaka,kobayashi1} suggesting
 that Fe$^{2+}$ ion is in the high spin configuration 
$t_{2g}^4e_g^2$ with  $S=2$.
As four  electrons occupy the threefold 
degenerate $t_{2g}$-levels, the ground state of Fe$^{2+}$  has a threefold
orbital degeneracy. This degeneracy, similarly to the spin one, can be described by pseudospin quantum numbers. 
The spin $S=2$ and pseudospin $\tau=1$ are thus the building blocks of an 
effective theory for insulating Sr$_2$FeWO$_6$.
In the double perovskite structure of Sr$_2$FeWO$_6$ Fe and W ions 
form a rock-salt structure. Each  type  of ions resides on a face-centered-cubic (fcc) sublattice. The fcc lattice is weakly frustrated and allows ordered
anisotropic AFM structures.  But weak frustration of the lattice alone can not be 
responsible for the very low magnetic transition temperature. 
The reason why $T_{N}$ is low, we believe, is as follows.
The superexchange interaction, that  couples magnetic moments of Fe ions,
is mediated by two types of diamagnetic ions, along the path Fe-O-W-O-Fe, 
with $90^{\circ}$ and $180^{\circ}$  angles between Fe and W ions. Because of this, the resulting Fe-Fe exchange 
is rather weak and the corresponding transition temperature is low.
The nearest-neighbor (nn) and next-nearest-neighbor (nnn) Fe moments
are coupled, respectively, by $90^{\circ}$ and  $180^{\circ}$ Fe-W-Fe bonds.
Depending on values and signs of the nn and nnn exchange energies,
different types of AFM structures can be stabilized on the fcc lattice.
To our knowledge no experimental determination of the kind of AFM order taking place in Sr$_2$FeWO$_6$
has ever been undertaken. 

In this paper we  derive  an effective  spin-orbital model from  the
Hubbard Hamiltonian of 
correlated  Fe-3d electrons hybridized with uncorrelated 5d electrons of W and determine the magnetic ground-state structure as a function of the model parameters.
We assume that oxygen degrees of freedom are integrated out and consider an effective
hybridization between Fe and W $d$-states.
Performing a fourth  order perturbation expansion in the hopping parameters 
around  the atomic limit we get  an  effective $S=2$ Hamiltonian
 for a system with six electrons in five $d$ orbitals.
It may be worthwhile to note that our derivation is based 
only on the geometrical structure and electronic configuration of the system. 
As such it can be applied to a variety of systems presenting the same 
characteristics: any rock-salt with a high-spin $t_{2g}^4e_g^2$
 configuration, like, e.g., FeO.

The paper is organized as follows: 
In the next section we introduce   the multi-band Hubbard Hamiltonian, 
perform the fourth order perturbation expansion and write the effective spin-orbital Hamiltonian in which the orbital degrees of freedom
are described  in terms of  the pseudospin  representation.
The derivation of the effective Hamiltonian demands a lot 
of algebra with rather cumbersome expressions.
To facilitate the reader, we have tried to use, whenever possible, 
a pictorial representation of states and processes, relegating both the fourth order perturbation scheme and all the final 
expressions in the Appendices A and B, correspondingly.
In section III  the classical ground state phase diagram of the effective 
spin-orbital model  is discussed, as a function of model parameters, 
within a variational procedure.
In the same section we also discuss the range of the external parameters 
appropriate to the case of Sr$_2$FeWO$_6$ and sketch some results.
In section IV we draw our conclusions.

\section{Effective Hamiltonian.}

Here we introduce the five-band Hubbard Hamiltonian,
 and perform a  perturbative expansion in the hopping parameters 
up to the fourth order 
to describe all  the low-lying excited states of the Fe-W system in terms of spin and orbital degrees of freedom 
of  iron sites, only.

\subsection{Ground state.}
We start from the
multi-band  Hubbard Hamiltonian: \begin{equation}
H=H_0+H_{\text{cf}}+H_{0W}+H'_t ~,
\label{acca}
\end{equation}
\noindent where $H_0$ is the on-site term 
of the form:
\begin{eqnarray}
H_{0}&=&U_{1}\sum_{i,\alpha}n_{i\alpha\uparrow}n_{i\alpha\downarrow}+
\frac{1}{2}(U_2-J)\sum_{i,\sigma, \alpha\neq \alpha '}
n_{i\alpha\sigma}n_{i\alpha'\sigma}  \nonumber \\
&+&U_2\sum_{i,\alpha\neq \alpha'}n_{i\alpha\uparrow}n_{i\alpha'\downarrow}+
J\sum_{i,\alpha\neq \alpha'} d_{i\alpha\uparrow}^{+}
d_{i\alpha\downarrow}^{+}d_{i\alpha'\downarrow}
d_{i\alpha'\uparrow} \nonumber \\
&-&J\sum_{i,\alpha\neq \alpha'} d_{i\alpha\uparrow}^{+}
d_{i\alpha\downarrow}d_{i\alpha'\downarrow}^{+}
d_{i\alpha'\uparrow} ~,
\label{h2}
\end{eqnarray}

\noindent $U_1$ and $U_2$ are the Coulomb repulsion among electrons in the same and in different orbitals, respectively, and
$J$ is the  Hund's coupling constant. Due to the cubic symmetry, the relation 
$U_1=U_2+2J$ holds. The annihilation and creation operators, 
$d_{i\sigma\alpha}$ and $d^{\dagger}_{i\sigma\alpha}$ 
refer to Fe-3d orbitals at site $i$, of 
type $\alpha$ (one of $x^2-y^2$, $3z^2-r^2 $, $xy$, $xz$ or $yz$) 
and with spin $\sigma=\uparrow,\downarrow$, 
$n_{i\alpha\sigma}=d^{\dagger}_{i\sigma\alpha}d_{i\sigma\alpha}$.

The octahedral crystal  field splitting 
of d-orbitals into a 
$e_g$-doublet ($x^2-y^2$ and $3z^2-r^2 $) and a lower-lying $t_{2g}$ triplet 
($xy$, $xz$ and $yz$)
 is taken into account for both Fe and W ions by the term:\cite{cf}
\begin{equation}
H_{cf} = \Delta_{e_g} \sum_{ii',\alpha=x^2-y^2,3z^2-r^2} 
(n_{i\alpha\sigma}+\tilde{n}_{i'\alpha\sigma})
\label{cf}
\end{equation}
\noindent where 
$\tilde{n}_{i'\alpha\sigma}$ counts the electrons on W sites $i'$. 
The summation is for all $i$ and $i'$ running over Fe and W sublattices,
 respectively.
Finally, the last term of the local part of the Hamiltonian, $H_{0W}$, represents the single particle energy level of the W ion with respect to Fe$^{(6)}$:
\begin{equation}
H_{0W} = \Delta_{\text{CT}} \sum_{i'\alpha\sigma} \tilde{n}_{i'\alpha\sigma}
~~,
\label{oW}
\end{equation}
where $ \Delta_{\text{CT}}=E(\text{Fe}^{(5)}\text{W}^{(1)})-E(\text{Fe}^{(6)}\text{W}^{(0)})$ is the charge transfer
energy, i.e., the minimum energy needed to
 remove one electron from Fe$^{(6)}$ ion and put it to an empty W$^{(0)}$
 site. Here and below an upper index denotes the number of $d$-electrons
 on a given ion. 
The kinetic term $H'_t$ is considered as a perturbation:
\begin{eqnarray}
H'_t\!&=&\!-\!\sum_{ii'\sigma\alpha\beta}t_{\alpha\beta}
\big( d_{i\sigma\alpha }^{\dagger}\tilde{d}_{i'\sigma\beta} + h.c. \big)
\label{Ht}
\end{eqnarray}
The summation is over nearest-neighbor Fe-W sites, and all possible
spin  and  orbital indices. Here $\tilde{d}^{\dagger}_{i'\sigma \alpha}$ 
($\tilde{d}_{i'\sigma \alpha}$) creates (annihilates) a 5d-electron on  W ions, 
with spin $\sigma$ in the orbital $\alpha$.
In a cubic lattice there is no hopping between $e_g$ and $t_{2g}$.
 The transfer  matrix  $t_{\alpha\beta}$ is 
diagonal in the $t_{2g}$ manifold 
for all the orbitals $xy$, $xz$, $yz$, 
with non-zero matrix elements $t$ only for the bonds laying 
in the corresponding plane (i.e., along $x$ direction, the hopping integral is zero for $yz$ orbitals and equal to $t$ for the other two).
 Due to the shape of the $e_g$
orbitals, their  hybridization is different in
the three cubic directions thus leading to direction dependent hopping
with the anisotropic transfer matrix elements  $t_{\alpha \beta}$
($\alpha = 3z^2-r^2 $ and $\beta =x^2-y^2$) given by:
\begin{eqnarray}
t^{x,y}_{\alpha \beta}=
t_e\left(\!\!\begin{array}{cc}
1/4 &\!
\mp\sqrt{3}/4\\
\mp\sqrt{3}/4&\!3/4
\end{array}
\!\!\right)~,\;\;\;
t^{z}_{\alpha \beta}=
t_e\left(\!\!\begin{array}{cc}
1 &\!\;\;
0\\
0&\!\;\;0
\end{array}
\!\!\right)
\label{2}
\end{eqnarray}

In constructing the effective Hamiltonian we  also 
assume that the  Hund's coupling $J$ is strong enough 
to win the competition with the
crystal field splitting $\Delta_{e_g}$.
This implies  that Fe$^{2+}$-ion  in the ground state  is in the high-spin
configuration $t_{2g}^4e_{g}^2 $ with total spin $S=2$, and
 justifies the exclusion from our zeroth order of the low-spin state, $t_{2g}^6e_{g}^0$ ($S=0$) and of the
intermediate-spin state, $t_{2g}^5e_{g}^1$ ($S=1$), both higher in energy.
We can write the ground state as a collection of atomic states $|\alpha_{j}\rangle$, each characterized by two quantum numbers: the spin $\vec{S}$ and  pseudospin $\vec{\tau}$. This latter describes the orbital occupation, namely the position of double occupied orbital in the $t_{2g}$-manifold. Since there are three possibilities to
put the double occupied orbital in the manifold, we can describe it by $\vec{\tau} =1$, with the same algebra of the usual spin operator.
The orbital quantization axis can be selected arbitrarily, and  we
choose the following convention which pictorially can be presented as:\\ 

\begin{picture}(50,30)
\put (10,0){\framebox(12,12){$\uparrow\downarrow$}}
\put (22.05,0){\framebox(12,12){$\uparrow$}}
\put (34.1,0){\framebox(12,12){$\downarrow$}}
\put (15.95,12.6){\framebox(12,12){$\uparrow$}}
\put (28.05,12.6){\framebox(12,12){$\uparrow$}}
\put (90,12){\makebox(50,0){$\rightarrow ~ |\alpha _1\rangle =|S_z=2,\tau_z=1\rangle$}}
\end{picture}\\

\begin{picture}(50,30)
\put (10,0){\framebox(12,12){$\uparrow$}}
\put (22.05,0){\framebox(12,12){$\uparrow\downarrow$}}
\put (34.1,0){\framebox(12,12){$\uparrow$}}
\put (15.95,12.6){\framebox(12,12){$\uparrow$}}
\put (28.05,12.6){\framebox(12,12){$\uparrow$}}
\put (90,12){\makebox(50,0){$\rightarrow ~ |\alpha _2\rangle =|S_z=2,\tau_z=0\rangle$}}
\end{picture}\\

\begin{picture}(50,30)
\put (10,0){\framebox(12,12){$\uparrow$}}
\put (22.05,0){\framebox(12,12){$\uparrow$}}
\put (34.1,0){\framebox(12,12){$\uparrow\downarrow$}}
\put (15.95,12.6){\framebox(12,12){$\uparrow$}}
\put (28.05,12.6){\framebox(12,12){$\uparrow$}}
\put (90,12){\makebox(50,0){$\rightarrow ~ |\alpha _3\rangle =|S_z=2,\tau_z=-1\rangle$}}
\end{picture}\\

In this notation, the upper boxes represent the $e_g$ orbital
doublet, $e_g^{(1)}=3z^2-r^2$ (left upper box) 
and $e_g^{(2)}=x^2-y^2$ (right upper box).
The lower boxes refer
to the three-fold degenerate $t_{2g}$ orbitals.
 with $t_{2g}^{(1)}=xy$ (left box), $t_{2g}^{(2)}=yz$ (central box) and
$t_{2g}^{(3)}=xz$ (right box).
The atomic ground state energy is
$E^{(6)}_{g}=15U_2-8J+2\Delta_{e_g}$. Because of the three-fold orbital and the five-fold spin degeneracies, the ground state is 15-fold degenerate.
Such a degeneracy is then lifted by the hopping term, $H'_t$,  as detailed 
below.

\subsection{Excited states and perturbation theory.}
In this subsection we briefly describe the perturbation scheme and classify the excited states
 according to their  total spin $S$, their $S_z$ projection and orbital configurations.  Since we are interested in describing the superexchange interactions between iron sites, we need to go beyond second order perturbation theory in the hopping Hamiltonian $H'_t$ and deal with fourth-order perturbative expansion. In fact, we neglect the direct Fe-Fe hopping  and consider only the hopping through the W sites.
The explicit form of $H'_t$ is such that the fourth order eigenvalue equation simplifies to (see Appendix A):
\begin{widetext}
\begin{eqnarray}
{\Bigl |}\sum_{mln}
\frac{\langle \alpha |H'_t|\beta_m \rangle
\langle \beta_m |H'_t|\beta_l \rangle
\langle \beta_l |H'_t|\beta_n \rangle
\langle \beta_n |H'_t|\alpha' \rangle}
{(E_{\alpha} - E_{\beta_m})(E_{\alpha} - E_{\beta_l})
(E_{\alpha} - E_{\beta_n})}- E {\Bigr |}=0 
\label{eqeigen}
\end{eqnarray}
\end{widetext}
where 
$|\alpha \rangle$ and  $|\alpha ' \rangle$ are two possible  
ground-states for $H_0$ (all the Fe$^{2+}$-ions in the $t_{2g}^4e_g^2$ configuration, all 5d W-orbitals empty), while
$|\beta_{m(l,n)}\rangle$ denotes all possible intermediate excited states, as imposed by $H'_t$. 
The first and third step states, $|\beta_{m(n)}\rangle$, must have five electrons on one Fe-ion and one electron in an adjacent W-ion 
(${\rm Fe^{(5)}}{\rm W^{(1)}}$), 
while the second-step state, $|\beta_{l}\rangle$, must be of the kind 
${\rm Fe^{(5)}}{\rm W^{(0)}}{\rm Fe^{(7)}}$ or 
${\rm Fe^{(5)}}{\rm W^{(2)}}{\rm Fe^{(5)}}$  on 
 the three sites involved in the hopping process (see Fig.~\ref{process}).
Our aim is to find out a representation of the effective Hamiltonian, 
leading to this same eigenvalue equation,
in terms of the spin and orbital degrees of freedom of iron only.
We  consider the full multiplet structure of the intermediate states with five and seven electrons which can be grouped according to their energies and the total value of the spin. Moreover, depending on the orbital occupation 
of Fe$^{(6)}$ ion in the ground state
and the orbital symmetry ($e_g$ or $t_{2g}$) of the hopping electron,
different  intermediate configurations can be reached.
As already stated, in the cubic symmetry the hopping is allowed only 
within $t_{2g}$-manifold and $e_g$-manifold, separately.

Consider the excited states which involve $t_{2g}$ hopping, first.
The intermediate atomic states with seven electrons Fe$^{(7)}_t$  have total spin $S=3/2$. The manifold is 12-fold degenerate (4(spin)$\times$3(orbital)) and one of the possible states can be depicted as:\\

\begin{picture}(50,30)
\put (10,12){\makebox(10,0){ $\text{Fe}^{(7)}_{~t} =$ }}
\put (40,0){\framebox(12,12){$\uparrow\downarrow$}}
\put (52.05,0){\framebox(12,12){$\uparrow\downarrow$}}
\put (64.1,0){\framebox(12,12){$\uparrow$}}
\put (45.95,12.6){\framebox(12,12){$\uparrow$}}
\put (58.05,12.6){\framebox(12,12){$\uparrow$}}
\end{picture}\\

\noindent The three-fold orbital degeneracy
 arises from the $t_{2g}$ unpaired electron. The energy of this state is 
$E^{(7)}_{t}=21U_2-7J+2\Delta $.

 The intermediate states with five electrons
split over a manifold with different values of total spin and 
energies.

The excited states with maximum spin configuration have  $S=5/2$ and no orbital degeneracy. The
6-fold degeneracy is entirely due to the spin. 
The state with maximum $S_z$ is:\\

\begin{picture}(50,30)
\put (10,12){\makebox(10,0){$\text{Fe}^{(5)}_{a} =$ }}
\put (40,0){\framebox(12,12){$\uparrow$}}
\put (52.05,0){\framebox(12,12){$\uparrow$}}
\put (64.1,0){\framebox(12,12){$\uparrow$}}
\put (45.95,12.6){\framebox(12,12){$\uparrow$}}
\put (58.05,12.6){\framebox(12,12){$\uparrow$}}
\end{picture}\\

\noindent The energy is
$E^{(5)}_{a}=10U_2-10J+2\Delta$, the minimum excitation
energy of $\text{Fe}^{(5)}$ states.

All the other $\text{Fe}^{(5)}$ states with all  orbitals singly occupied 
have total spin $S=3/2$ and can be pictured as:

\begin{picture}(50,30)
\put (20,12){\makebox(20,0){  $\text{Fe}^{(5)}_{b_1} =
\frac{1}{\sqrt{2}} {\bigg (}$ }}
\put (70,0){\framebox(12,12){$\uparrow$}}
\put (82.05,0){\framebox(12,12){$\uparrow$}}
\put (94.1,0){\framebox(12,12){$\uparrow$}}
\put (75.95,12.6){\framebox(12,12){$\uparrow$}}
\put (88.05,12.6){\framebox(12,12){$\downarrow$}}
\put (118,12){\makebox(3,0){$ - $}}
\put (130,0){\framebox(12,12){$\uparrow$}}
\put (142.05,0){\framebox(12,12){$\uparrow$}}
\put (154.1,0){\framebox(12,12){$\uparrow$}}
\put (135.95,12.6){\framebox(12,12){$\downarrow$}}
\put (148.05,12.6){\framebox(12,12){$\uparrow$}}
\put (173,12){\makebox(5,0){$ \bigg ) $}}
\end{picture}\\

\begin{picture}(50,30)
\put (20,12){\makebox(20,0){  $\text{Fe}^{(5)}_{b_2} =
\frac{1}{\sqrt{2}} {\bigg (}$ }}
\put (70,0){\framebox(12,12){$\uparrow$}}
\put (82.05,0){\framebox(12,12){$\uparrow$}}
\put (94.1,0){\framebox(12,12){$\downarrow$}}
\put (75.95,12.6){\framebox(12,12){$\uparrow$}}
\put (88.05,12.6){\framebox(12,12){$\uparrow$}}
\put (118,12){\makebox(3,0){$ - $}}
\put (130,0){\framebox(12,12){$\downarrow$}}
\put (142.05,0){\framebox(12,12){$\uparrow$}}
\put (154.1,0){\framebox(12,12){$\uparrow$}}
\put (135.95,12.6){\framebox(12,12){$\uparrow$}}
\put (148.05,12.6){\framebox(12,12){$\uparrow$}}
\put (173,12){\makebox(5,0){$ \bigg ) $}}
\end{picture}\\

\hspace{0.3cm}
\begin{picture}(50,30)
\put (10,12){\makebox(20,0){ $\text{Fe}^{(5)}_{b_3}
=\frac{1}{\sqrt{6}} {\bigg (2}$ }}
\put (60,0){\framebox(12,12){$\uparrow$}}
\put (72.05,0){\framebox(12,12){$\downarrow$}}
\put (84.1,0){\framebox(12,12){$\uparrow$}}
\put (65.95,12.6){\framebox(12,12){$\uparrow$}}
\put (78.05,12.6){\framebox(12,12){$\uparrow$}}
\put (102,12){\makebox(3,0){$ - $}}
\put (110,0){\framebox(12,12){$\uparrow$}}
\put (122.05,0){\framebox(12,12){$\uparrow$}}
\put (134.1,0){\framebox(12,12){$\uparrow$}}
\put (115.95,12.6){\framebox(12,12){$\uparrow$}}
\put (128.05,12.6){\framebox(12,12){$\downarrow$}}
\put (151,12){\makebox(3,0){$ - $}}
\put (160,0){\framebox(12,12){$\uparrow$}}
\put (172.05,0){\framebox(12,12){$\uparrow$}}
\put (184.1,0){\framebox(12,12){$\uparrow$}}
\put (165.95,12.6){\framebox(12,12){$\downarrow$}}
\put (178.05,12.6){\framebox(12,12){$\uparrow$}}
\put (200,12){\makebox(5,0){${\bigg )}$}}
\end{picture}\\

\hspace{0.3cm}
\begin{picture}(50,30)
\put (18,12){\makebox(20,0){ $\text{Fe}^{(5)}_{b_4} =\frac{2}{3} \sqrt{\frac{3}{10}} {\bigg (}$ }}
\put (70,0){\framebox(12,12){$\uparrow$}}
\put (82.05,0){\framebox(12,12){$\downarrow$}}
\put (94.1,0){\framebox(12,12){$\uparrow$}}
\put (75.95,12.6){\framebox(12,12){$\uparrow$}}
\put (88.05,12.6){\framebox(12,12){$\uparrow$}}
\put (112,12){\makebox(3,0){$ - $}}
\put (120,0){\framebox(12,12){$\uparrow$}}
\put (132.05,0){\framebox(12,12){$\uparrow$}}
\put (144.1,0){\framebox(12,12){$\uparrow$}}
\put (125.95,12.6){\framebox(12,12){$\uparrow$}}
\put (138.05,12.6){\framebox(12,12){$\downarrow$}}
\put (161,12){\makebox(3,0){$ - $}}
\put (170,0){\framebox(12,12){$\uparrow$}}
\put (182.05,0){\framebox(12,12){$\uparrow$}}
\put (194.1,0){\framebox(12,12){$\uparrow$}}
\put (175.95,12.6){\framebox(12,12){$\downarrow$}}
\put (188.05,12.6){\framebox(12,12){$\uparrow$}}
\put (211,12){\makebox(5,0){${\bigg )}$}}
\end{picture}\\

\hspace{0.3cm}
\begin{picture}(50,30)
\put (50,12){\makebox(5,0){$ -\sqrt{\frac{3}{10}} \bigg ( $}}
\put (75,0){\framebox(12,12){$\uparrow$}}
\put (87.05,0){\framebox(12,12){$\uparrow$}}
\put (99.1,0){\framebox(12,12){$\downarrow$}}
\put (80.95,12.6){\framebox(12,12){$\uparrow$}}
\put (93.05,12.6){\framebox(12,12){$\uparrow$}}
\put (118,12){\makebox(3,0){$ + $}}
\put (130,0){\framebox(12,12){$\downarrow$}}
\put (142.05,0){\framebox(12,12){$\uparrow$}}
\put (154.1,0){\framebox(12,12){$\uparrow$}}
\put (135.95,12.6){\framebox(12,12){$\uparrow$}}
\put (148.05,12.6){\framebox(12,12){$\uparrow$}}
\put (173,12){\makebox(5,0){$ \bigg ) $}}
\end{picture}\\

Each of these states is 4-fold degenerate because of the spin (only the
states with  maximum $S_z$ are shown). They give the 
same contribution to the effective Hamiltonian and below, for convenience,
we denote them with the common notation $\text{Fe}^{(5)}_{b}$.
Their energy is $E^{(5)}_{b}=10U_2-5J+2\Delta$.

Finally, there are the states with one unoccupied orbital.
They  have total spin $S=3/2$ and the global degeneracy is
 24 (4(spin)$\times$6(orbital)).\\

\begin{picture}(50,30)
\put (20,12){\makebox(20,0){ $\text{Fe}^{(5)}_{c_{\pm}}
=\frac{1}{\sqrt{2}} {\bigg (}$ }}
\put (70,0){\framebox(12,12){$\uparrow$}}
\put (82.05,0){\framebox(12,12){$\uparrow\downarrow$}}
\put (94.1,0){\framebox(12,12){}}
\put (75.95,12.6){\framebox(12,12){$\uparrow$}}
\put (88.05,12.6){\framebox(12,12){$\uparrow$}}
\put (118,12){\makebox(3,0){$ \pm $}}
\put (130,0){\framebox(12,12){$\uparrow$}}
\put (142.05,0){\framebox(12,12){}}
\put (154.1,0){\framebox(12,12){$\uparrow\downarrow$}}
\put (135.95,12.6){\framebox(12,12){$\uparrow$}}
\put (148.05,12.6){\framebox(12,12){$\uparrow$}}
\put (173,12){\makebox(5,0){$ \bigg ) $}}
\end{picture}\\

 The  energies of these states are 
$E^{(5)}_{c_+}=10U_2-5J+2\Delta $ and 
$E^{(5)}_{c_-}=10U_2-3J+2\Delta $. 

Now we consider the intermediate states which can be reached by $e_g$ hopping.
Those with seven electrons have spin $S=3/2$ and can be presented as 
the following :\\

\hspace{0.3cm}
\begin{picture}(50,30)
\put (10,12){\makebox(10,0){$\text{Fe}^{(7)}_{e} =$ }}
\put (40,0){\framebox(12,12){$\uparrow$}}
\put (52.05,0){\framebox(12,12){$\uparrow\downarrow$}}
\put (64.1,0){\framebox(12,12){$\uparrow$}}
\put (45.95,12.6){\framebox(12,12){$\uparrow\downarrow$}}
\put (58.05,12.6){\framebox(12,12){$\uparrow$}}
\end{picture}\\

Their degeneracy is 4(spin)$\times$3(orbital $t_{2g}$)$\times$2(orbital $e_g$)=24. 
Their energy is: $E^{(7)}_{e}=21U_2-7J+3\Delta $. 

The states with five electrons arising from the 
$e_g$ hopping have also total spin $S=3/2$ and can be drawn as\\

\hspace{0.2cm}
\begin{picture}(50,30)
\put (20,12){\makebox(20,0){$\text{Fe}^{(5)}_{d_{\pm}} =\frac{1}{\sqrt{N_{\pm}}} {\bigg (}$ }}
\put (70,0){\framebox(12,12){$\uparrow$}}
\put (82.05,0){\framebox(12,12){$\uparrow\downarrow$}}
\put (94.1,0){\framebox(12,12){$\uparrow$}}
\put (75.95,12.6){\framebox(12,12){$\uparrow$}}
\put (88.05,12.6){\framebox(12,12){}}
\put (118,12){\makebox(3,0){$ \pm \alpha $}}
\put (130,0){\framebox(12,12){$\uparrow$}}
\put (142.05,0){\framebox(12,12){}}
\put (154.1,0){\framebox(12,12){$\uparrow$}}
\put (135.95,12.6){\framebox(12,12){$\uparrow$}}
\put (148.05,12.6){\framebox(12,12){$\uparrow\downarrow$}}
\put (173,12){\makebox(5,0){$ \bigg ) $}}
\end{picture}\\

\noindent where the normalization factors are defined  as $N_{\pm}=\frac{J^2+\Delta^2_{\pm}}{\Delta^2_{\pm}}$, $\alpha = \frac{1}{\sqrt{2}}\sqrt{1+\frac{\Delta }{\sqrt{J^2+\Delta^2}}}$, and
$\Delta_{\pm}=\Delta \pm\sqrt{J^2+\Delta^2}$.
These states have energy
$E^{(5)}_{d_{\pm}}=10U_2-4J+2\Delta \mp \sqrt{J^2+\Delta^2}$ and
their degeneracy is 4(spin)$\times$3(position of doubly occupied orbital)$\times$2(position of hole)=24.  


\subsection{Classification of virtual processes and derivation of the spin-orbital Hamiltonian.}

In order to derive the effective spin-orbital Hamiltonian in the simplest way
it is useful to classify all the allowed hopping processes depending on the specific geometry  of the fcc lattice.
It becomes then possible to select the orbital dependence for each process and for a given projection of the spin coupling (whether FM or AFM) along the direction $ij$ of the bond.
The general form for the Hamiltonian turns out to be:
\begin{eqnarray}
H_{\rm eff} =  
\sum_{ij} -{\big [}
\vec S_i\cdot \vec S_j +6 {\big ]}O^{(F)}_{ij}+
{\big [}\vec S_i\cdot \vec S_j-4 {\big ]}O^{(A)}_{ij}
\label{spinorb}
\end{eqnarray}
\noindent where  $ij$ denote the summation over both nn and nnn  Fe-ions of
the fcc lattice.
The terms  $O^{(F(A))}_{ij}$ are the orbital, energy-dependent, contributions to the effective Hamiltonian for FM or AFM bonds $ij$, respectively,  which are explicitly given in Appendix B. The two spin-dependent terms are projectors over $\vec{S}_i+\vec{S}_j=4$ (FM coupling) and  $\vec{S}_i+\vec{S}_j=0$ (AFM coupling), respectively.
Below we clarify the procedure adopted to derive the spin-orbital Hamiltonian (\ref{spinorb}). We try to avoid as much as possible formulas  and present the derivation in a pictorial way.
 Following Fig.~\ref{process} we consider four types of processes (from $A$ 
to $D$), according to the way the intermediate states can be reached.

Depending on the kind and number of ions involved in the intermediate steps, it is possible to distinguish the following processes:
\begin{eqnarray*}
&A -\text{processes}: \\
&{\rm Fe^{(6)}}{\rm W^{(0)}}{\rm Fe^{(6)}}\rightarrow
{\rm Fe^{(5)}}{\rm W^{(1)}}{\rm Fe^{(6)}}\rightarrow
{\rm Fe^{(5)}}{\rm W^{(0)}}{\rm Fe^{(7)}} \\
&\rightarrow{\rm Fe^{(5)}}{\rm W^{(1)}}{\rm Fe^{(6)}}\rightarrow
{\rm Fe^{(6)}}{\rm W^{(0)}}{\rm Fe^{(6)}}
\end{eqnarray*}
\begin{eqnarray*}
&B/C -\text{processes}:\\
&{\rm Fe^{(6)}}{\rm W^{(0)}}{\rm Fe^{(6)}}\rightarrow
{\rm Fe^{(5)}}{\rm W^{(1)}}{\rm Fe^{(6)}}\rightarrow
{\rm Fe^{(5)}}{\rm W^{(2)}}{\rm Fe^{(5)}}\\
&\rightarrow{\rm Fe^{(5)}}{\rm W^{(1)}}{\rm Fe}^{(6)}\rightarrow
{\rm Fe^{(6)}}{\rm W^{(0)}}{\rm Fe^{(6)}}
\end{eqnarray*}
\begin{eqnarray*}
&D -\text{processes}: \\
&{\rm Fe^{(6)}}{\rm W^{(0)}}{\rm W^{(0)}}{\rm Fe^{(6)}}\rightarrow
{\rm Fe^{(5)}}{\rm W^{(1)}}{\rm W^{(0)}}{\rm Fe^{(6)}}\\
&\rightarrow{\rm Fe^{(5)}}{\rm W^{(1)}}{\rm W^{(1)}}{\rm Fe^{(5)}} 
\rightarrow
{\rm Fe^{{(6)}}}{\rm W^{(0)}}{\rm W^{(1)}}{\rm Fe^{{(5)}}}\\
&\rightarrow
{\rm Fe^{(6)}}{\rm W^{(0)}}{\rm W^{(0)}}{\rm Fe^{(6)}}
\end{eqnarray*}
\noindent 
where the upper index denotes  the number of electrons on each ion. 
A first basic distinction is easily drawn between the $A$-process and all 
the others. The former involve the formation of a Fe$^{(7)}$ 
state at the second step, while none of the latter type processes does.
 
\begin{figure}
\epsfysize=40mm
\centerline{\epsffile{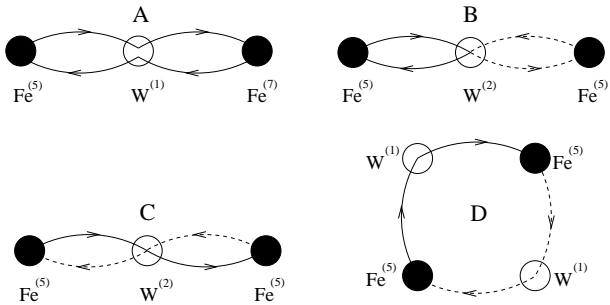}}
\vspace{0.5cm}
\caption{Different hopping processes: dashed and full lines correspond to different electrons. $A$- and $B$- processes exist for both 90$^\circ$ and 180$^\circ$ Fe-W-Fe angles (only 180$^\circ$ are shown). $C$-processes exist only for the 180$^\circ$-angle and $D$-processes only for the 90$^\circ$-angle.}
\label{process}
\end{figure}

On the contrary, all  $B$-, $C$- and $D$-processes are characterized by a Fe$^{(5)}$ state for
both iron ions involved in a virtual process. The difference among them lies in the way these states are 
reached. For $B$- and $C$-processes there is only one intermediate W ion with two electrons on it, differently from the the $D$-process, where two distinct W 
ions are involved. Finally, the difference between $B$- and $C$-processes lies in 
the behavior of the two virtual electrons. In the first case they go back to the original Fe-ion where they came from. On the contrary for C-processes there is an 
exchange of the two electrons, as shown in Fig.~\ref{process}, and 
each of them moves toward the other Fe ion.
For the sake of clarity it should be reminded that we are considering 
the specific case of the fcc sublattice of Fe ions in Sr$_2$FeWO$_6$: this allows some simplifications in the derivation of the effective spin-orbital Hamiltonian, 
$H_{\text{eff}}$. In particular, it is possible to build the following table, where we list the contributions to $H_{\text{eff}}$ depending on the 
characteristics of each single term.\\
\begin{ruledtabular}
\begin{tabular}{lcccc}
          &  $A$   & $B$  & $C$  &   $D$  \\ \hline
$aa$        & {\rm FM } &  {\rm AFM (T1)} & {\rm  FM (T2,nnn) } & {\rm  FM (T1,nn) }     \\ 
$ bb$        & {\rm AFM} &  {\rm AFM (T1)} &  {\rm FM (T2,nnn)}  & {\rm  FM (T1,nn) }       \\ 
$cc$        & {\rm AFM} &  {\rm AFM (T1)} & {\rm  FM (T2,nnn) } & {\rm  FM (T1,nn) }   \\ 
$dd$        & {\rm AFM} &  {\rm AFM (T1)} & {\rm  FM (T2,nnn)}  &  {\rm FM (T1,nn) }    \\ 
$ac$       & --   &  {\rm FM (T1)} &  {\rm AFM (T2,nnn)} &  {\rm AFM (T1,nn) }     \\ 
$bc$       & --  &  {\rm AFM (T1)} & {\rm  FM (T2,nnn)}  &  {\rm FM (T1,nn) }      \\ 
$ab$       & --  &  {\rm FM  (T1)} &  {\rm AFM (T2,nnn)} & {\rm  AFM (T1,nn)}       \\ 
\end{tabular}
\end{ruledtabular}\mbox{}\\
Each column in the table represents one of the four possible processes drawn in Fig.~\ref{process}. 
Each row represents the states of the Fe$^{(5)}$-ion reached at the second step of the virtual process.
For example, for $B$-, $C$- and $D$-processes, $aa$ means that the virtual intermediate state is made of both Fe$^{(5)}_{a}$, while $ac$ indicates one Fe$^{(5)}_{a}$-state and one Fe$^{(5)}_{c\pm}$-state. 
 In the case of the A process, only one Fe$^{(5)}$-state is present 
and ``$aa$'' refers to this single Fe$^{(5)}_{a}$-state, while $ac$ has no meaning.
The table gives us the magnetic character of the bond $ij$ involved in 
the process (whether FM or AFM), 
if it requires a nn or a nnn contribution, and  its orbital signature.
 In this respect T$1$ means that 
only one type of orbital is allowed in the process (ie, a contribution proportional to $t_{xy}^4$), while 
T$2$  means that there must be two different orbitals (for example a 
contribution proportional to 
$t_{xy}^2t_{xz}^2$). If there are no orbital or neighbors 
specifications,
 it means that both kinds 
are allowed.
With this in mind, the derivation of the orbital part reported in Appendix B 
follows quite straightforward.

As an example, we show how to derive the term (\ref{oAcc}) (see Appendix B), corresponding to the
A-process towards states $c_{\pm}$. In order to have a state $c_{\pm}$
on a Fe$^{(5)}$-ion, the electron that jumps to the W-site and then forms the
Fe$^{(7)}_t$ state, must belong to one of the singly occupied t$_{2g}$-states of the original Fe$^{(6)}$-ion.
Only in this case, in fact, the Fe$^{(5)}$-ion is left with one doubly
occupied, one singly occupied and one vacant $t_{2g}$-state. The electron
is free to jump on W site, that is empty, and it has some restriction to hop 
to
the next Fe$^{(6)}$ ion, because of the Pauli principle. Only some jumps
are allowed and they do depend on the relative orbital occupations of the
two Fe$^{(6)}$-ions involved in the process.
It is worthwhile to note that, if the spins of the two Fe$^{(6)}$-ions
involved were ferromagnetically coupled, no hopping processes could have
taken place because of the Pauli principle: the hopping electron would
find all the singly occupied orbitals of the final Fe$^{(6)}$-ion filled with the same spin
projection. Thus, superexchange between 
c$_{\pm}$ states can take place
only when the two Fe$^{(6)}$ ions are coupled antiferromagnetically, as
shown in the previous table.

Let us now analyze the orbital dependence. For simplicity we consider two
specific configurations, of those that we called T1 and T2 in the table, e.g., the term
depending on t$_{xy}^4$ and that depending on
t$_{xy}^2$t$_{xz}^2$. The first hopping process is allowed only when 
$\tau_{iz}\not= 1$ and $\tau_{jz}\not=1$, i.e. $xy$-orbital states are
singly occupied at both $i$ and $j$ Fe-sites involved.
The second is active only when both
$\tau_{iz}$ and $\tau_{jz}$ are either 1 or -1, i.e. either $xy$- or 
$xz$-orbital states are doubly occupied at both sites. During this process
both orbital quantum numbers are either reduced or increased  by two, 
respectively. In the case of a doubly occupied $xy$-orbital, 
the $xz$-electron hops leaving a hole in this orbital state. The resulting state
is not an eigenstate of the Hamiltonian (\ref{h2})
 and is then projected over an eigenstate  $c_{\pm}$ given by
the superposition of  states with $xy$- and  $xz$-holes. Therefore, 
an electron from the doubly occupied $xy$-orbital 
of Fe$^{(7)}$ can hop back to Fe$^{(5)}$ ion. 
This changes the initial orbital
configuration $\tau_{iz}=\tau_{jz}=1$ to the final  
$\tau_{iz}=\tau_{jz}=-1$ form.
In terms of the pseudospin quantum numbers, the two contributions are:
\begin{eqnarray*}
t_{xy}^4
\Bigl[(1-\tau_{iz}^2)+\frac{1}{2}\tau_{iz}(\tau_{iz}-1)\Bigr]
\Bigl[(1-\tau_{jz}^2)+
\frac{1}{2}\tau_{jz}(\tau_{jz}-1)\Bigr]\\
\pm\frac{1}{4}t_{xy}^2t_{xz}^2
\Bigl[\tau_{i}^+\tau_{i}^+\tau_{j}^+\tau_{j}^+ + 
\tau_{i}^-\tau_{i}^-\tau_{j}^-\tau_{j}^-\Bigr]~,
\end{eqnarray*} 
where $\pm$ depends on which of the  c$_{\pm}$ states are considered.
These terms enter into 
Eq. (\ref{eqeigen}) with a weight taking into account both
spin (through Clebsch-Gordan coefficients) and orbital projection of the single
Slater determinant into the correlated state c$_{\pm}$. 
The weight, or normalization coefficient (see the Table in Appendix B), of a given
process is obtained by equating the matrix elements of the Hamiltonian (\ref{spinorb})
to the corresponding ones obtained directly using the perturbation theory
(\ref{eqeigen}). To calculate the weight of the proccess under consideration,
for simplicity, we conider diagonal matrix elements of 
$H_{\text{eff}}$ in the spin subspace over the bond configuration 
$S_i^z=2$ and $S_j^z=-2$.
The $SU(2)$ spin symmetry of the problem guarantees that consideration of the
nondiagonal matrix elements in the spin subspace, i. e. processes that involve
spin flips, will lead to  the same result.
For the present process this weight is $\gamma_{A}^{cc} =
\frac{1}{16}=\frac{1}{8}\big(\frac{1}{\sqrt{2}}\big)^2$, where the factor $1/8$ 
normalizes $|\langle\vec S_i\cdot \vec S_j-4\rangle|=8$ 
for the bond we consider and the remaining $\big(\frac{1}{\sqrt{2}}\big)^2$ is due to the projection of the single Slater determinant into the states c$_{\pm}$. The square power is due to the fact that the
 virtual state,
not eigen-state of the local Hamiltonian, is projected twice into the
corresponding eigenstate in Eq. (\ref{eqeigen}).

All the other contributions, detailed in Appendix B, have been found in a
similar way.

\section{Ground state phase diagram.}

In this section we discuss the classical ground state phase diagram of the 
spin-orbital model, Eq. (\ref{spinorb}).  We factorize spin and orbital degrees of freedom and
perform a mean-filed type analysis of the possible ordered phases. 
The factorization is justified because of the two different energy scales
governing these degrees of freedom: 
spins are coupled by virtual processes that involve  both 
$e_g$ and $t_{2g}$ electron transfers, 
while only the latter electrons generate 
the coupling between the orbital degrees of freedom, as $e_g$ orbitals are half-filled.
The ratio of spin and orbital exchange energies thus scales approximately 
as $t_{e}^4/t^4$ 
and is much larger than unity in the situation of interest.
Therefore, it is reasonable to average out spin degrees of freedom, first, by
considering all possible ordered magnetic structures on a fcc
 centered cubic lattice
and then for each type of magnetic ordering find an orbital configuration, that minimizes the total energy by means of a variational procedure.

We consider the following magnetic structures that are 
possible ordered states for a nn and a nnn Heisenberg Hamiltonian on a fcc lattice\cite{smart}: 

{\bf i)} Ferromagnetic state (FM);

{\bf ii)} Type-I antiferromagnetic (AFM-I) ordering, which consists of 
ferromagnetic \{110\} planes stacked antiferromagnetically in the third direction;

\begin{figure}
\epsfysize=45mm
\centerline{\epsffile{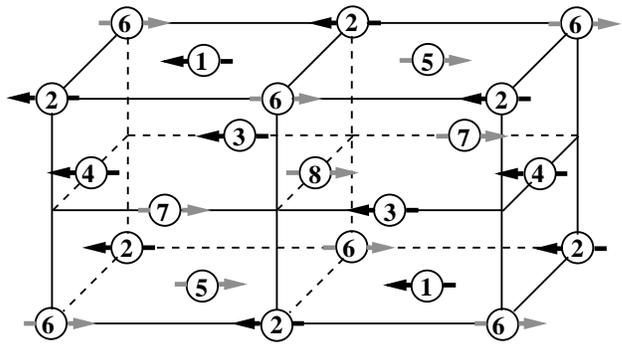}}
\vspace{0.5cm}
\caption{Type-II antiferromagnetic ordering of Fe moments on the face centered
cubic lattice.}
\label{afm2}
\end{figure}

{\bf iii)} Type-II antiferromagnetic (AFM-II) structure, consisting of
ferromagnetic \{111\} planes, each  being antiferromagnetically coupled to 
adjacent layers. This magnetic structure is presented
 in Fig.~\ref{afm2}, where only the Fe atoms are shown.

\subsection{Minimization procedure.}

The trial wave function can be written in general as follows:
\begin{equation}
|\Psi\rangle =\Pi_{i}~|\Psi_i\rangle =\Pi_{i}~|\psi^o_i\rangle |\phi^s_i\rangle
\label{variational}
\end{equation} 
where $|\psi^o_i\rangle$ and $|\phi^s_i\rangle$
 refer to  orbital and spin states at site $i$ , respectively.
Since $H_{\rm eff}$ factors  into an orbital $H^o_{\rm eff}$ and a spin $H^s_{\rm eff}$ part and its action involves two sites at a time, its average value over the above state takes the form:
\begin{eqnarray}
\begin{array}{c}
\langle\Psi_i|\langle\Psi_j|H_{\rm eff}|\Psi_j\rangle |\Psi_i\rangle =
\\[0.2cm]
\langle\psi^o_i|\langle\psi^o_j|H^o_{\rm eff}|\psi^o_j\rangle |\psi^o_i\rangle
\times
\langle\phi^s_i|\langle\phi^s_j|H^s_{\rm eff}|\phi^s_j\rangle |\phi^s_i\rangle
\end{array}
\label{av}
\end{eqnarray}

The orbital part of the variational wave function can be written as 
\begin{eqnarray}
\parallel\psi^o_i\rangle
=\cos\theta_i|0\rangle_{i}+\sin\theta_i
(\cos\psi_i|1\rangle_{i}+\sin\psi_i|-1\rangle_{i}) ~.
\end{eqnarray}
allowing all three orbital states $|\tau_z=0\rangle_{i}$, $|\tau_z=1\rangle_{i}$ and $|\tau_z=-1\rangle_{i}$  with a relative weight determined by the minimization procedure with respect to the variational parameters $\theta_i$ and $\psi_i$. The expectation values of $H^o_{\rm eff}$ over the trial orbital wave function
for  both ferromagnetic  and antiferromagnetic bonds
are then expressed in terms of variational parameters $\theta_i$ 
and  $\psi_i$ and are quite lengthy.
On the contrary,
the average of spin variables is straightforward 
in a mean field treatment. For a ferromagnetic bond
$\langle \vec{S}_i\cdot\vec{S}_j+6\rangle=10$ and 
for an antiferromagnetic bond 
$\langle \vec{S}_i\cdot\vec{S}_j-4\rangle=-8$ in the classical ground state.

The energy per Fe atom,  $E^{\text{Fe}}$,  is given by the sum of the energy contributions from all the bonds contained in the cell, divided by the number of atoms (each Fe-ion has 12 nn and 6 nnn).
AFM-II phase  has the biggest  cell, with eight Fe atoms (see Fig.~\ref{afm2}), and  for generality we consider it as the unit cell for both  FM and AFM-I phases, too. Hence, we have sixteen  variational parameters
($\theta_{1-8}$ and $\psi_{1-8}$ for the eight sites of the unit cell). 
We have then performed a numerical minimization of the ground state energy
with respect to all the sixteen independent angular variational angles $\theta_i$ and $\psi_i$. By taking the absolute minimum we  determine the orbital configuration
that minimizes the total energy of the system for 
a given magnetic order. The ground state magnetic and orbital structure 
are determined by selecting the  state with the lowest energy.
The results  are presented below.

\subsection{Model parameters}

In order to proceed further with the ground state phase
diagram of $H_{\text{eff}}$, we fix below  the values of the parameters
that have already been estimated in the literature and 
we try to give a reasonable guess for the remainings.
\begin{itemize}
\item Crystal field splitting $\Delta_{e_g}$:  The value of the $e_{g}$ -- $t_{2g}$
splitting $\Delta_{e_g}\simeq1.24$ eV has been estimated for Fe$^{2+}$ ion in Ref. 
\onlinecite{anderson} and we take this value in our analysis.
\item Exchange parameter $J$: The exchange part of Coulomb energy is known to
  be little screened in solids. It can thus be estimated trough the atomic
  values of Racah parameters B and C ($J=5/2B+C$). Given the values of $B$ and
  $C$ from Ref. \onlinecite{bocquet}, one obtains $J\simeq0.78$ eV.
\item Coulomb energy $U_2$:  Intra-atomic Coulomb repulsion is generally
  considerably screened in solids and  can not be simply related to its 
atomic value. The estimates reported in the literature suggest that
$U_2=5-7$ eV (Refs. \onlinecite{anderson,bocquet,takahashi}), however 
we also consider variation of $U_2$ in a wider range.
\item Transfer integrals $t$ and $t_e$: We fix, in most of the cases, the
 ratio of $e_{g}$ and $t_{2g}$ 
transfer integrals to be  $t_e/t=3$ as suggested by the Muffin-Tin Orbital theory.\cite{harrison} We are not aware of any estimates of absolute values of the transfer integrals between Fe and W ions. However $t_{2g}$ hopping amplitude has been estimated to be $t\simeq 0.25$ eV for Fe-Re bond.\cite{millis}        
We may consider the same value of $t_{2g}$ transfer for  Fe-W bond too, since
for both Re and W ions this transfer involves  $5d$-shell.
\item Charge-transfer gap $\Delta_{\text{CT}}$: We can estimate a lower boundary for the charge transfer gap between the high spin state of Fe$^{(6)}$ and W ions 
in two possible ways as follows. 
First, the difference in $\Delta_{\text{CT}}$ for Re and W based systems
can be approximately taken equal to the energy mismatch of  $5d$-levels of
 these ions: $\Delta_{\text{CT}}(\text{W})-\Delta_{\text{CT}}(\text{Re})=
E^{(5d)}(\text{W})-E^{(5d)}(\text{Re})$.
The latter mismatch is of the order of $\sim 1.4$ eV.\cite{harrison}
A lower estimate of $\Delta_{\text{CT}}(\text{W})$ is then obtained by 
putting $\Delta_{\text{CT}}(\text{Re})=0$ for the metallic Re based compound.
We get $\Delta_{\text{CT}}>1.4$ eV for the W based compound. 
Second, the insulating behavior of the system 
implies that the charge-transfer gap is larger than the bandwidth of those electrons that are involved in the 
lowest energy charge-transfer.\cite{zaanen,aligia}
Taking the above reported value for $t_{2g}$ transfer one obtains the even more strict constraint
$\Delta_{\text{CT}}>2$ eV.    
\end{itemize}  

\subsection{Results and discussion}

\begin{figure}
\epsfysize=60mm
\centerline{\epsffile{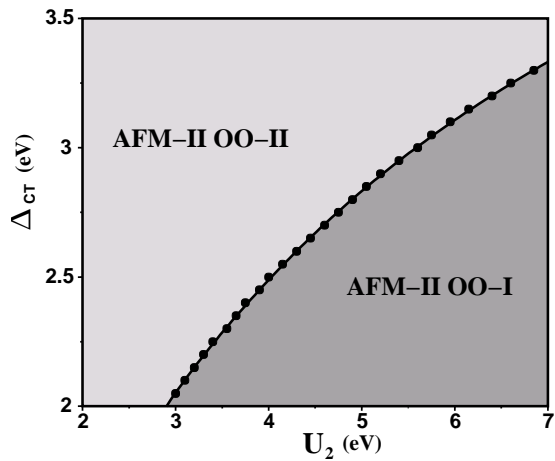}}
\vspace{0.5cm}
\caption{Ground state classical phase diagram of the spin-orbital model 
(\ref{spinorb}) in $\Delta_{\text{CT}}$ vs $U_2$ parameter plane. Here AFM-II denotes 
type-II antiferromagnetic ordering, OO-I and OO-II correspond to two  types of orbital ordering described in the text.}
\label{oopd}
\end{figure}

In Fig. \ref{oopd} we show the ground state phase diagram of  
the spin-orbital model (\ref{spinorb}) in $\Delta_{\text{CT}}-U_2$ plane. We fix the other parameters as 
discussed above and take $t_e/t=3$. 
Because of this ratio,  $e_g$ superexchange, which is antiferromagnetic, 
overwhelms the superexchange due to the $t_{2g}$ hopping.
Therefore the ground state is always antiferromagnetic and it turns out to be of type-II.
 
Given this magnetic structure,
 orbitals  order in such a way to further reduce the ground state energy. 
The type of orbital ordering depends on whether 
$2\Delta_{\text{CT}} \agt U_2$ (orbital ordering of type-II, OO-II) 
or  
$2\Delta_{\text{CT}} \alt U_2$ (orbital ordering of type-I, OO-I). 
In the former case virtual processes with
singly occupied W-states (A-processes) become dominant. Instead in the latter case it is energetically more favorable to create doubly occupied W states rather than to pay a Coulomb energy at Fe sites (B-, C-, D-processes). 
In Table \ref{oo} we list the types of orbital orderings together with the corresponding 
structure.  The numbers from 1 to 8 
label  the eight atoms of the unit cell shown in Fig.~\ref{afm2}.
The ket-vectors denote the type of doubly occupied $t_{2g}$ 
orbital at each atom for each orbital ordering.  

To understand the orbital structure of the AFM-II state it is convenient to divide
the fcc lattice of Fe ions, in four simple cubic sublattices.
These four sublattices are generated by the pairs of nonequivalent atoms
of the unit cell: (1,5), (2,6), (3,7), and (4,8) (see Fig.~\ref{afm2}).
Each sublattice is AFM and the moments are coupled only by
nnn exchange, while the coupling between sublattices is due to the nn
exchange. If we denote by  $J_1$  and  $J_2$ the nn and nnn exchange constants
of Fe spins, respectively, and consider only spin degrees of freedom, then the
 ground state energy in the AFM-II
 phase depends only on $J_2$, because
there is an equal number of nn ferro-
and antiferro-bonds. Therefore the contributions from $J_1$ to the 
 ground state energy are  canceled out.
 However, in the present case, due to the orbital part of the effective
 Hamiltonian (\ref{spinorb}), inter-sublattice couplings (i.e.  nn
 spin-orbital couplings) also contribute to the ground state energy. 
The antiferromagnetic superexchange for 
a given sublattice is maximized when $t_{2g}$ orbitals are ordered ferromagnetically
and all equally populated. Switching on the inter-sublattice coupling
 does not affect the ferromagnetic-type orbital order within each sublattices. However, the type of occupied orbital state changes, in order to minimize the contribution of nn energy to the ground state.
As a result, there are two types of orbitally ordered ground states (OO-I and OO-II), depending on the parameters (see Fig.~\ref{oopd}).
The structure of these orbital states is explicitly given in Table~\ref{oo}.

\begin{table}
\caption{\label{oo}Types of orbital ordering with corresponding structures}
\begin{ruledtabular}
\begin{tabular}{lccc}
&  {\rm OO-I}   & {\rm OO-II}  & {\rm OO-III}    \\ \hline
{\rm 1}        &$|yz\rangle$ &$(|yz\rangle+|xz\rangle)/\sqrt{2}$& $|xz\rangle$      \\ 
{\rm 2}        &$|xy\rangle$ &$|xy\rangle$ &$|xy\rangle$          \\ 

{\rm 3}        &$|xz\rangle$ &$ |xz\rangle$  & $|xz\rangle$  \\ 
{\rm 4}        &$(|yz\rangle+|xz\rangle)/\sqrt{2}$ &
$(|yz\rangle+|xy\rangle)/\sqrt{2}$   &$|xy\rangle$      \\ 
{\rm 5}        &$|yz\rangle$           &$(|yz\rangle+|xz\rangle)/\sqrt{2}$   & $|yz\rangle$         \\ 
{\rm 6}        & $|xy\rangle$          &$|xy\rangle$   &$|yz\rangle$   \\ 
{\rm 7}        & $|xz\rangle$          & $|xz\rangle$   &  $|xz\rangle$        \\ 
{\rm 8}        & $(|yz\rangle+|xz\rangle)/\sqrt{2}$&$(|yz\rangle+|xz\rangle)/\sqrt{2}$   &  $|xy\rangle$ 
\end{tabular}
\end{ruledtabular}
\end{table}

We have also evaluated $J_{1}$ and $J_{2}$  
 averaged over the 
bond directions. The transition temperature $T_{N}$
can be directly linked to the nnn exchange energy $J_2$ by the formula:
 $T_{N}=2S(S+1)zJ_{2}/3$ where $z=6$ is a number of next-nearest neighbors.
For $S=2$ and the experimental values of $T_{N}\simeq 16-37$ K one obtains
$J_{2}\simeq 0.06-0.15$ meV. Fixing $t_e/t=3$, $U_2=5$ eV and varying 
$\Delta_{\text{CT}}$ in the range $2-7$ eV we obtain the following range of 
$J_2$: $J_2=0.03-0.14$ meV for $t=0.20$ eV and $J_2=0.08-0.35$ meV for 
$t=0.25$ eV. The exchange energy scales as the forth power of the
 transfer integrals and is thus very sensitive to the variation of $t$, nonetheless
we can conclude that the superexchange theory gives the correct order of magnitude of the experimental transition temperature.
It is worth mentioning that $t_{2g}$ electrons contribute approximately only to
one-tenth of the value of $J_2$. Our analysis also suggests that  $J_1$ is
 antiferromagnetic as well, but around half of $J_2$. 
Next-nearest neighbor spins are thus coupled stronger than nearest-neighbor ones. 

\begin{figure}
\epsfysize=60mm
\centerline{\epsffile{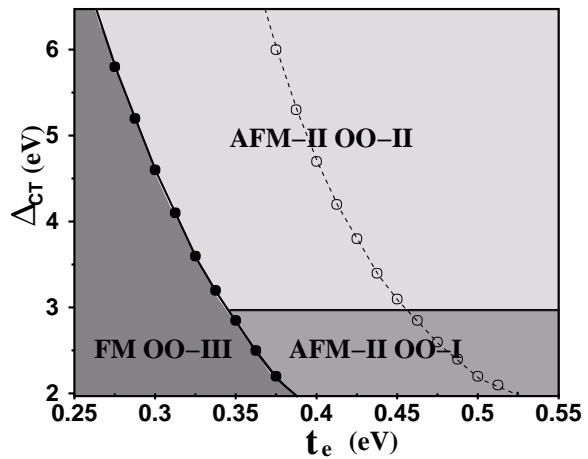}}
\vspace{0.5cm}
\caption{Same as in Fig.~\ref{oopd} in $\Delta_{\text{CT}}$ vs $t_e$ parameter plane. Here FM denotes ferromagnetic phase and OO-III corresponds to the orbital ordering described in the text. Filled and open circles stand for the 
FM--AFM-II phase boundaries for $U_2=5$ and $2.5$ eV, respectively.}
\label{mopd}
\end{figure}

To complete the discussion on the ground state properties of the 
spin-orbital Hamiltonian (\ref{spinorb}), in the following we relax
the constraint $t_e/t=3$, fixing $t=0.25$ eV and $U_2=5$ eV, and consider the
phase diagram in $t_e$ and $\Delta_{\text{CT}}$ parameter plane. The result
is presented in Fig. \ref{mopd}. For small values of $e_g$ hopping, a ferromagnetic phase is stabilized. Its orbital ordering  
(OO-III) is shown in Table \ref{oo}.
In this configuration the system maximizes the nn ferromagnetic exchange, 
$J_1>0$.
The exchange between nnn neighbors is always antiferromagnetic, $J_2<0$,   and increases
with $t_e$. At some critical value of the $e_{g}$ transfer, at which $J_1+J_2=0$,
the FM ordering becomes unstable and  a transition to the AFM-II state takes place 
(see filled circles in Fig. \ref{mopd}). For smaller values of the Coulomb energy, $U_2$,
the phase space for ferromagnetic ordering is enlarged and the transition to the AFM-II
state occurs for larger values of $t_{e}$. The FM--AFM-II phase boundary for 
$U_2=2.5$ eV is shown as open circles in Fig. \ref{mopd}.

\section{Summary}

To summarize, we have developed a superexchange theory
for insulating double-perovskite compounds such as Sr$_2$FeWO$_6$.
The superexchange interaction, that  couples the magnetic moments of Fe$^{2+}$ ions through the diamagnetic W$^{6+}$ ions, takes place on Fe-W-Fe 
bonds with $90^{\circ}$ and $180^{\circ}$  angles. The theory has been 
implemented in an effective spin-orbital Hamiltonian formulated
in terms of spin ($S=2$) and orbital pseudo-spin ($\tau=1$) degrees of freedom
of the iron ion only, arising from the threefold degeneracy of $t_{2g}$-levels.
We have studied magnetically and orbitally ordered states and 
constructed the classical ground state phase diagram of the effective Hamiltonian.
We have shown that for realistic values of the model parameters the ground
state is antiferromagnetic, as experimentally observed. Our results suggests,
that the antiferromagnetic order is of type-II, i.e.,  $\{111\}$
ferromagnetic planes stacked antiferromagnetically. This prediction can be tested by means of polarized neutron experiments.  We have evaluated the exchange
energies for these magnetic structure and showed that the theory 
gives an estimate of the transition temperature which is compatible with the 
experimental value. The magnetic structure is accompanied by an orbital ordering
in which each cubic-sublattice of iron ions shows a ferro-type orbital ordering. The type of occupied orbital states changes from one sublattice to the other.
The energy scale of orbital degrees of freedom is one order of magnitude smaller than the spin one. The is due to the fact that orbitals are coupled
by $t_{2g}$ electrons, only,  while both $e_{g}$ 
(with larger transfer amplitude)  and  $t_{2g}$ virtual processes lift the spin
degeneracy. 
Finally, we note that in the present paper we have just dealt with the
electronic mechanisms to lift the ground state orbital degeneracy.

However, there are two other possible
 mechanisms, not discussed here, that can lift the orbital degeneracy,
such as the lattice distortion, or Jahn-Teller (JT) effect, and the spin-orbit coupling.
The latter breaks the rotational spin invariance and opens a spin gap: Therefore a correct order of magnitude for this effect could be obtained by the experimental measurement of the spin-wave gap.
As for the lattice mechanism, it should support the orbital
 ordering that minimizes the electronic energy of the system
and thus enlarge the stability range of the orbital order, because
the JT effect in this compound is mainly governed by the physics of isolated clusters rather 
than by cooperative effects. In fact, the oxygen octahedra surrounding Fe ions are independent one another, as they do not share any common oxygen-ion. 
Therefore  JT distortions are not correlated as far as change 
in the electronic energy is concerned: Distortions of the octahedra around W ions
with empty $5d$-level do not cause any considerable 
change in the electronic energy.

In conclusion, if the electronic mechanism dominates the other two, then the orbital order can be stabilized only 
at very low temperatures, much lower than the
magnetic transition temperature and in between the system is  magnetically ordered but orbitally disordered.  
In order to clarify which of the above mechanism is dominant further experimental studies
of the magnetic and structural properties of the system are necessary.

\acknowledgments

We would like to thank  C. R. Natoli for a critical reading of the manuscript.
G. J. acknowledges kind hospitality at the theory group of the LNF-INFN, 
Frascati. 

\appendix
\begin{widetext}
\section{Derivation of the fourth order energy correction.}
Consider the Hamiltonian:
\begin{eqnarray}
H=H_0+\lambda H'
\end{eqnarray}
where $H_0$ is exactly solvable ($H_0 \psi_0 =E_0 \psi_0$) and 
$\lambda =$ is some ``small'' non-dimensional parameter.
Following, for example Ref.\onlinecite{bransden}, it is straightforward to write the coupled equations to get eigenvalue and eigenvector 
corrections to $E_0$ and $\psi_0$ up to the fourth order in $\lambda$:
\begin{eqnarray*}
(H_0-E_{0})\psi_{0}=0~, 
(H'-E_{1})\psi_{0}+(H_0-E_{0})\psi_{1}=0~,
(H'-E_{1})\psi_{1}+(H_0-E_{0})\psi_{2}-E_{2}\psi_{0}=0~,\\
(H'-E_{1})\psi_{2}+(H_0-E_{0})\psi_{3}
-E_{2}\psi_{1}-E_{3}\psi_{0}=0~,\\
(H'-E_{1})\psi_{3}+(H_0-E_{0})\psi_{4}-
E_{2}\psi_{2}-E_{3}\psi_{1}-E_{4}\psi_{0}=0~.
\end{eqnarray*}
Multiplying scalarly all equations by
$(\psi_{0})^*$ and solving with respect to 
 the fourth order energy correction, $E_4$,  we get:

\begin{eqnarray}
E_{4}&=&\sum_{l\neq k} \sum_{m\neq k}\bigg( \sum_{n\neq k}
\frac{H'_{kl}H'_{lm}H'_{mn}H'_{nk}}
{(E_k-E_l)(E_k-E_m)(E_k-E_n)}-\sum_{n\neq k}
\frac{H'_{kl}H'_{lm}\delta_{mn}H'_{nk}H'_{kk}}
{(E_k-E_l)(E_k-E_m)(E_k-E_n)}\\\nonumber 
&-&\frac{H'_{kl}H'_{lk}H'_{km}H'_{mk}+
H'_{kl}H'_{lm}H'_{mk}H'_{kk}-
H'_{kl}\delta_{lm}H'_{mk}H'^2_{kk}}
{(E_k-E_m)(E_k-E_l)^2}\bigg)
\end{eqnarray}
where we use notation: $H'_{km}=\langle \psi_k |H'| \psi_m\rangle $.
Given the form of $H'_t$, Eq. (\ref{Ht}), it follows that $H'_{kk}=0$, ie, the second,
 fourth and fifth terms do not contribute. Also the third term can be
 neglected, because it does not give rise to an effective exchange between
 Fe-sites, being the product of two second-order processes. 
This justifies our formula (\ref{eqeigen}) in Sec.II.A.
\section{Orbital terms and energy denominators.}
Here we write explicitly the orbital contributions 
to the effective Hamiltonian (\ref{spinorb}).

First we consider virtual processes with  $t_{2g}$-electrons involved. 
All orbital contributions $O_{ij}^{(F(A))}$ for a bond $ij$ 
can be written in the following general form:
\begin{eqnarray}
O_{ij}^{(F(A))} = \sum_{P,ll'} O_{ij,P}^{ll'}~,~\text{where}~
O_{ij,P}^{ll'}=\frac{\gamma_{P}^{ll'}}
{D_{P}^{ll'}}\tilde{O}_{ij,P}^{ll'}
\end{eqnarray}
and $D_{P}^{ll'}$ denotes an energy denominator, $\tilde{O}_{ij,p}^{ll'}$ 
is an orbital contribution expressed by means of the 
pseudospin operators, $\gamma_{P}^{ll'}$ are normalization factors, which 
are given in the Table at the end of the Appendix B, $P$ 
denotes one of the four virtual 
hopping processes $A$, $B$, $C$, and  $D$, depicted in Fig.~\ref{process},
 and $l$ and $l'$ are the kind of excited 
states involved in the hopping process (i.e. $a$, $b$, and $c$ introduced
in Sec. II.B).

In the following,  we list all the terms, classified for the kind of process.

{\bf $A$-process:}
For processes involving excited states of  $a$- and   $b$- type  
the orbital contributions differ only because of the  energy denominators:
\begin{eqnarray}
D_{A}^{aa(bb)}=
\Delta_{\text{CT},a(b)}^2
\Bigl[E^{(7)}_{t}+E^{(5)}_{a(b)}-2E^{(6)}_{g}\Bigr]~,
\end{eqnarray}
here and below $\Delta_{\text{CT},l}$ denotes 
energy needed  for an electron transfer from  Fe$^{(6)}$ to W$^{(0)}$ 
when iron ion is left in a Fe$^{(5)}_{l}$ state. This energy is given by
$\Delta_{\text{CT},l}=\Delta_{\text{CT}}+E^{(5)}_{l}-E^{(5)}_{a}$.
The state $a$ is the lowest energy configuration of  Fe$^{(5)}$ and 
$\Delta_{\text{CT}}=\Delta_{\text{CT},a}$ is the lowest energy necessary for the
$\text{Fe}^{(6)}\rightarrow \text{W}^{(0)}$ charge transfer.

The corresponding pseudospin part is:
\begin{eqnarray}
\tilde{O}_{ij,A}^{aa(bb)}=
\frac{t_{xy}^4}{2}\tau_{iz}(\tau_{iz}+1)\Bigl[\frac{1}{2}\tau_{jz}(\tau_{jz}-1)+
(1-\tau_{jz}^2)\Bigr]+
\frac{t_{xz}^4}{2}\tau_{iz}(\tau_{iz}-1)\Bigl[\frac{1}{2}\tau_{jz}
(\tau_{jz}+1)+
(1-\tau_{jz}^2)\Bigr]+
t_{yz}^4(1-\tau_{iz}^2)\tau_{jz}^2
\nonumber
\\
+\frac{t_{xy}^2t_{xz}^2}{4}
\Bigl[\tau_{i}^-\tau_{i}^-\tau_{j}^+\tau_{j}^+ + \tau_{i}^+\tau_{i}^+\tau_{j}^-\tau_{j}^-\Bigr]
+
\frac{t_{xy}^2t_{yz}^2}{2}
\Bigl[\tau_{i}^-\tau_{iz}\tau_{jz}\tau_{j}^+ + \tau_{iz}\tau_{i}^+\tau_{j}^-\tau_{jz}\Bigr]
+
\frac{t_{xz}^2 t_{yz}^2}{2}
\Bigl[\tau_{iz}\tau_{i}^-\tau_{j}^+\tau_{jz} + \tau_{i}^+\tau_{iz}\tau_{jz}\tau_{j}^-\Bigr]\nonumber\\
\label{oAaa}
\end{eqnarray}

If the excited state is 
Fe$^{(5)}_{c\pm}$ then the purely orbital 
contribution can be written as:
\begin{eqnarray}
\tilde{O}_{ij,A}^{cc}\!\!\!\!&=&
t_{xy}^4
\Bigl[(1-\tau_{iz}^2)+\frac{1}{2}\tau_{iz}(\tau_{iz}-1)\Bigr]
\Bigl[(1-\tau_{jz}^2)+
\frac{1}{2}\tau_{jz}(\tau_{jz}-1)\Bigr]\label{oAcc}\\
\!\!\!\!&+&\!\!\!\!t_{xz}^4
\Bigl[(1-\tau_{iz}^2)+\frac{1}{2}\tau_{iz}(\tau_{iz}+1)\Bigr]
\Bigl[(1-\tau_{jz}^2)+\frac{1}{2}\tau_{jz}(\tau_{jz}+1)\Bigr]
+t_{yz}^4\tau_{iz}^2\tau_{jz}^2\nonumber\\
\!\!\!\!&\pm&\!\!\!\!
\frac{t_{xy}^2t_{xz}^2}{4}
\Bigl[\tau_{i}^+\tau_{i}^+\tau_{j}^+\tau_{j}^+ + \tau_{i}^-\tau_{i}^-\tau_{j}^-\tau_{j}^-\Bigr]\pm
\frac{t_{xy}^2t_{yz}^2}{2}
\Bigl[\tau_{iz}\tau_{i}^+\tau_{jz}\tau_{j}^+ + \tau_{i}^-\tau_{iz}\tau_{j}^-\tau_{jz}\Bigr]\pm
\frac{t_{xz}^2 t_{yz}^2}{2}
\Bigl[\tau_{iz}\tau_{i}^-\tau_{jz}\tau_{j}^- + \tau_{i}^+\tau_{iz}\tau_{j}^+\tau_{jz}\Bigr]\nonumber
\end{eqnarray}
where $c$ can be either $c_+$ or $c_-$, and respectively, either  upper or
lower sign should be taken in Eq.~\ref{oAcc}. 
Corresponding  energy denominator is given by:
\begin{eqnarray}
D_{A}^{cc}=
\Delta_{\text{CT},c}^2
\Bigl[E^{(7)}_{t}+E^{(5)}_{c}-2E^{(6)}_{g}\Bigr]~.
\end{eqnarray}

{\bf $B$-process:} It  has only diagonal orbital contributions, 
as shown in the Table of section II. 
If $aa$-, $bb$-, $ab$- excited states are involved
then the pure orbital part is the same, and 
 can be written as:
\begin{eqnarray}
\tilde{O}_{ij,B}^{ll'}=
\frac{t_{xy}^4}{4}\tau_{iz}(\tau_{iz}+1)\tau_{jz}(\tau_{jz}+1)
+
\frac{t_{xz}^4}{4}\tau_{iz}(\tau_{iz}-1)\tau_{jz}(\tau_{jz}-1)
+
t_{yz}^4(1-\tau_{iz}^2)(1-\tau_{jz}^2)
\label{oBll}
\end{eqnarray}
For $cc$-hopping processes the orbital contribution  
$\tilde{O}_{ij,B}^{cc}$ 
is equal to diagonal part of $\tilde{O}_{ij,A}^{cc}$ and
reads as
\begin{eqnarray}
\tilde{O}_{ij,B}^{cc}&=&
t_{xy}^4
\Bigl[(1-\tau_{iz}^2)+\frac{1}{2}\tau_{iz}(\tau_{iz}-1)\Bigr]
\Bigl[(1-\tau_{jz}^2)+
\frac{1}{2}\tau_{jz}(\tau_{jz}-1)\Bigr]
\label{oBcc}\\
&+&
t_{xz}^4
\Bigl[(1-\tau_{iz}^2)+\frac{1}{2}\tau_{iz}(\tau_{iz}+1)\Bigr]
\Bigl[(1-\tau_{jz}^2)+\frac{1}{2}\tau_{jz}(\tau_{jz}+1)\Bigr]
+
t_{yz}^4\tau_{iz}^2\tau_{jz}^2
\nonumber
\end{eqnarray}
As for  $ac$- and  $bc$- processes, their orbital part   
are equal to the diagonal part of $\tilde{O}_{ij,A}^{aa(bb)}$ (\ref{oAaa})
given by
\begin{eqnarray}
\tilde{O}_{ij,B}^{ac(bc)}=
\frac{t_{xy}^4}{2}\tau_{iz}(\tau_{iz}+1)\Bigl[\frac{1}{2}\tau_{jz}(\tau_{jz}-1)+
(1-\tau_{jz}^2)\Bigr]+
\frac{t_{xz}^4}{2}\tau_{iz}(\tau_{iz}-1)\Bigl[\frac{1}{2}\tau_{jz}(\tau_{jz}+1)+
(1-\tau_{jz}^2)\Bigr]+
t_{yz}^4(1-\tau_{iz}^2)\tau_{jz}^2\nonumber\\
\label{oBac(bc)}
\end{eqnarray}
The general form for B-processes energy denominators is
\begin{eqnarray}
D_{B}^{ll'}=\Delta_{\text{CT},l}\Delta_{\text{CT},l'}\Bigl[
\Delta_{\text{CT},l}+\Delta_{\text{CT},l'}\Bigr]
\label{DB}
\end{eqnarray} 

{\bf $C$-process:} In this process two electrons with the same spin directions are
exchanged between iron sites. The intermediate step involves the formation of a doubly occupied 
W site. This implies that two different 
orbitals must participate to the hopping process because of Pauli principle.
The orbital contributions are thus  only non-diagonal.  
They are the same for all 
 processes of the kind $aa$, $bb$, $cc$, and $ab$: 

\begin{eqnarray}
\tilde{O}_{ij,C}^{ll'}&=&\pm
\frac{t_{xy}^2t_{xz}^2}{4}
\Bigl[\tau_{i}^-\tau_{i}^-\tau_{j}^+\tau_{j}^+ + \tau_{i}^+\tau_{i}^+\tau_{j}^-\tau_{j}^-\Bigr]\nonumber\\
&\pm&\frac{t_{xy}^2t_{yz}^2}{2}\Bigl[
\tau_{i}^-\tau_{iz}\tau_{jz}\tau_{j}^+ + \tau_{iz}\tau_{i}^+\tau_{j}^-\tau_{jz}\Bigr]\pm
\frac{t_{xz}^2t_{yz}^2}{2}
\Bigl[\tau_{iz}\tau_{i}^-\tau_{j}^+\tau_{jz} + \tau_{i}^+\tau_{iz}\tau_{jz}\tau_{j}^-\Bigr]~,
\label{oC}
\end{eqnarray}
where the lower sign should  be taken only when $l=c_-$ and $l'=c_+$. These processes differ one 
another only by the energy denominators.

For $ac$- and $bc$- processes we have
\begin{eqnarray}
\tilde{O}_{ij,C}^{ac(bc)}&=&
\pm
\frac{t_{xy}^2t_{xz}^2}{4}
(\tau_{i}^+\tau_{i}^+\tau_{j}^+\tau_{j}^+ +
\tau_{i}^-\tau_{i}^-\tau_{j}^-\tau_{j}^-)\nonumber\\
&\pm&
\frac{t_{xy}^2t_{yz}^2}{2}
(\tau_{iz}\tau_{i}^+\tau_{jz}\tau_{j}^+ + \tau_{i}^-\tau_{iz}\tau_{j}^-\tau_{jz})\pm
\frac{t_{xz}^2 t_{yz}^2}{2}
(\tau_{iz}\tau_{i}^-\tau_{jz}\tau_{j}^- + \tau_{i}^+\tau_{iz}\tau_{j}^+\tau_{jz})
\label{oCac}
\end{eqnarray}
where upper sign $(+)$ and lower sign $(-)$  correspond to $c=c_+$ and 
 $c=c_-$, respectively.

We note that along $90^\circ$ Fe-W-Fe bond (nn Fe ions)
 hopping is allowed only for one kind of orbital.
Therefore, being orbitally nondiagonal, 
$C$-processes are effective only for $180^\circ$ bond (nnn Fe ions), 
as shown in the Table of Sec. II.C. 
Energy denominators  for all orbital contributions have the following common form:
\begin{eqnarray}
D_{C}^{ll'}=\Delta_{\text{CT},l}\Delta_{\text{CT},l'}\Bigl[
\Delta_{\text{CT},l}+\Delta_{\text{CT},l'}\Bigr]~.
\end{eqnarray}

{\bf $D$-process:} This type of processes is effective only for $90^\circ$ Fe-W-Fe bonds
by definition [see Fig.~\ref{process}] and thus   can only be orbitally 
diagonal. All orbital contributions and energy denominators are 
equal to the corresponding ones of B-processes [see Eqs.~\ref{oBll}-\ref{DB}].

Now we consider  $e_g$  hopping processes.
The contributions 
to the Hamiltonian  from such processes does not depend on the orbital degree of freedom:
\begin{eqnarray}
O_{ij,P}^{dd}=
\frac{\gamma_{P}^{dd}}{D_{P}^{dd}}t_e^4~.
\end{eqnarray}
Here the extra factors, which depend on the type of process, and 
appear due to the anisotropic form of $e_g$ transfer (\ref{2}), are included
in normalization factors  $\gamma_{P}^{dd}$ given in the Table below.

   The energy denominators for different processes can be written in the
following forms. For A-process
\begin{eqnarray}
D_{A}^{dd}=
\Delta_{\text{CT},d}^2
\Bigl[E^{(7)}_{e}+E^{(5)}_{d}-2E^{(6)}_{g}\Bigr]~,
\end{eqnarray}
and for B-, C- and D-processes
\begin{eqnarray}
D_{C}^{ll'}=\Delta_{\text{CT},l}\Delta_{\text{CT},l'}\Bigl[
\Delta_{\text{CT},l}+\Delta_{\text{CT},l'}\Bigr]~.
\end{eqnarray}
where $d$ and $d'$  can be either $d_+$ or $d_-$ excited states.

In the following Table we list 
all normalization coefficients $\gamma_P^{ll'}$:\\

\begin{ruledtabular}
\begin{tabular}{lccccccc}
        &$aa$   & $bb$    & $cc$    & $dd$      &$ac$
        &bc &   ab\\\hline
 $A$  & $1/10$&$1/10$ &$1/16$&$1/8N_{\pm}$  & --           & --& --\\
 $B$          & $1/8$ &$2/25$ &$1/32$&$1/(8N_{\pm}^2)$&$1/20$&  $1/20$& $2/25$ \\
 $C$          & $1/10$&$8/125$&$1/40$&$1/(10N_{\pm}^2)$&$1/16$&$1/25$& $1/10$\\      
 $D$          & $1/10$&$8/25$ &$1/40$&$1/(10N_{\pm}^2)$&$1/16$&$1/25$& $1/10$\\
\end{tabular}
\end{ruledtabular}
\mbox{}\\
\end{widetext}



\begin{thebibliography}{99}
\bibitem{cf} We consider the same value of octahedral crystal field splitting 
for Fe and W ions, in order to reduce number of parameters.
\end{thebibliography}
\end{document}